\newcommand{\rhoe}{\rho^-e^+\nu} 
 
\newcommand{\kste}{K^{*-}e^+\nu} 
\newcommand{\ke}{K^-e^+\nu} 
\newcommand{\pie}{\pi^-e^+\nu} 
\newcommand{\drhoe}{D^0\to \rho^-e^+\nu} 
\newcommand{\dkste}{D^0\to K^{*-}e^+\nu} 
\newcommand{\dke}{D^0\to K^-e^+\nu} 
\newcommand{\dpie}{D^0\to \pi^-e^+\nu}

\newcommand{\vcs}{V_{cs}} 

\newcommand{\vcd}{V_{cd}} 
\newcommand{\vub}{V_{ub}}

\documentclass{ws-ijmpa}

\begin{document}

\markboth{F. Liu}
{Measurements of Absolute Branching Fractions for ...}  

%
\catchline{}{}{}{}{}
%

\title{Measurements of Absolute Branching Fractions for Exclusive $D^0$ Semileptonic Decays} 

\author{\footnotesize F. Liu \\ Representing the CLEO Collaboration}

\address{Southern Methodist U. Group, Wilson Lab, Cornell University, Ithaca, New York 14853}

\maketitle


\begin{abstract}
  Based on a data sample of 60 pb$^{-1}$ collected at the 
  $\psi(3770)$ resonance with the CLEO-c detector at CESR, 
  we present improved measurements of absolute branching 
  fractions for exclusive $D^0$ semileptonic decays 
  into $\ke$, $\pie$ and $\kste$, and the first observation
  and measurement of $D^{0} \to \rhoe$. 
  The determinations of the CKM matrix elements 
  $\vcs$ and $\vcd$ are reviewed. 

\end{abstract}

\section{Introduction} 
In the Standard Model (SM), the three generation quark mixing is described by 
the unitary Cabibbo-Kobayashi-Maskawa (CKM) matrix~\cite{CKM}. 
It governs all flavor transitions of quarks and $CP$
symmetry violation due to a complex phase of it. 
Precision determinations of the CKM matrix 
elements from weak decays of the relevant quarks 
provide powerful tests of the Standard Model. 

Charm meson ($c\bar q$) semileptonic decays allow the 
measurements  of the CKM matrix elements $\vcs$ and $\vcd$. 
The differential decay rate for exclusive 
semileptonic decays $D\to Pe^+\nu$ ($P$ stands for a 
pseudoscalar meson) with the electron mass neglected can be 
expressed as~\cite{detector}: 
\begin{eqnarray}
\frac{d\Gamma}{dq^2}=\frac{G_F^2}{24\pi^3}\left|V_{cq'}\right|^2p_P^3
\left|f_+(q^2)\right|^2, 
\label{diffrate} 
\end{eqnarray}   
where $G_F$ is the Fermi coupling constant, $q^2$ is the 
four-momentum transfer squared between the parent $D$ meson 
and the final state meson, $p_P$ is the momentum 
of the pseudoscalar meson in the $D$ rest frame, and $V_{cq'}$ is 
the relevant CKM matrix element, either $\vcs$ or $\vcd$. 
$f_+(q^2)$ is the form factor that measures the probability that 
the flavor changed quark ($q'$) from  $c\to q'\ell\nu$ transitions 
and the spectator anti-quark ($\bar q$) 
will form a meson in the final state. 

To determine $\vcs$ and $\vcd$, we need theoretical input of the absolute 
normalization of the form factor describing the decay at a 
fixed $q^2$ point, usually at $q^2=0$. $D$ semileptonic decays 
can also offer tests of the theoretical predictions of the form factors.  
Since the charm meson semileptonic decay form factors are closely related 
to those of $B$ meson semileptonic decays by heavy quark symmetry,  
precise knowledge of the charm semileptonic form factors will in turn 
help improve the determination of the CKM matrix element $\vub$.

\section{Analysis Technique and Event Selection } 

The data sample of 60 pb$^{-1}$ used for this analysis was collected at 
the $\psi(3770)$ resonance with the CLEO-c detector~\cite{detector} 
at the Cornell Electron Storage Ring (CESR).

In this analysis, we take advantage of the unique kinematics of 
the $D\bar D$ threshold production at the $\psi(3770)$ resonance which
provides very powerful means to 
reject backgrounds from misidentified and missing particles. 
We fully reconstruct one $D^0$ meson of the produced $D^0\bar D^0$ 
pairs 
 as a tag. We identify 
an electron and a set of hadrons recoiling against the tag, 
reconstructing the missing momentum and missing energy. 
The difference between the missing energy and missing momentum in a 
event $U=E_{miss}-p_{miss}$ will peak at zero if the event is 
correctly reconstructed due to the undetected neutrino. 
\begin{figure}[hbtp]
\centerline{\psfig{file=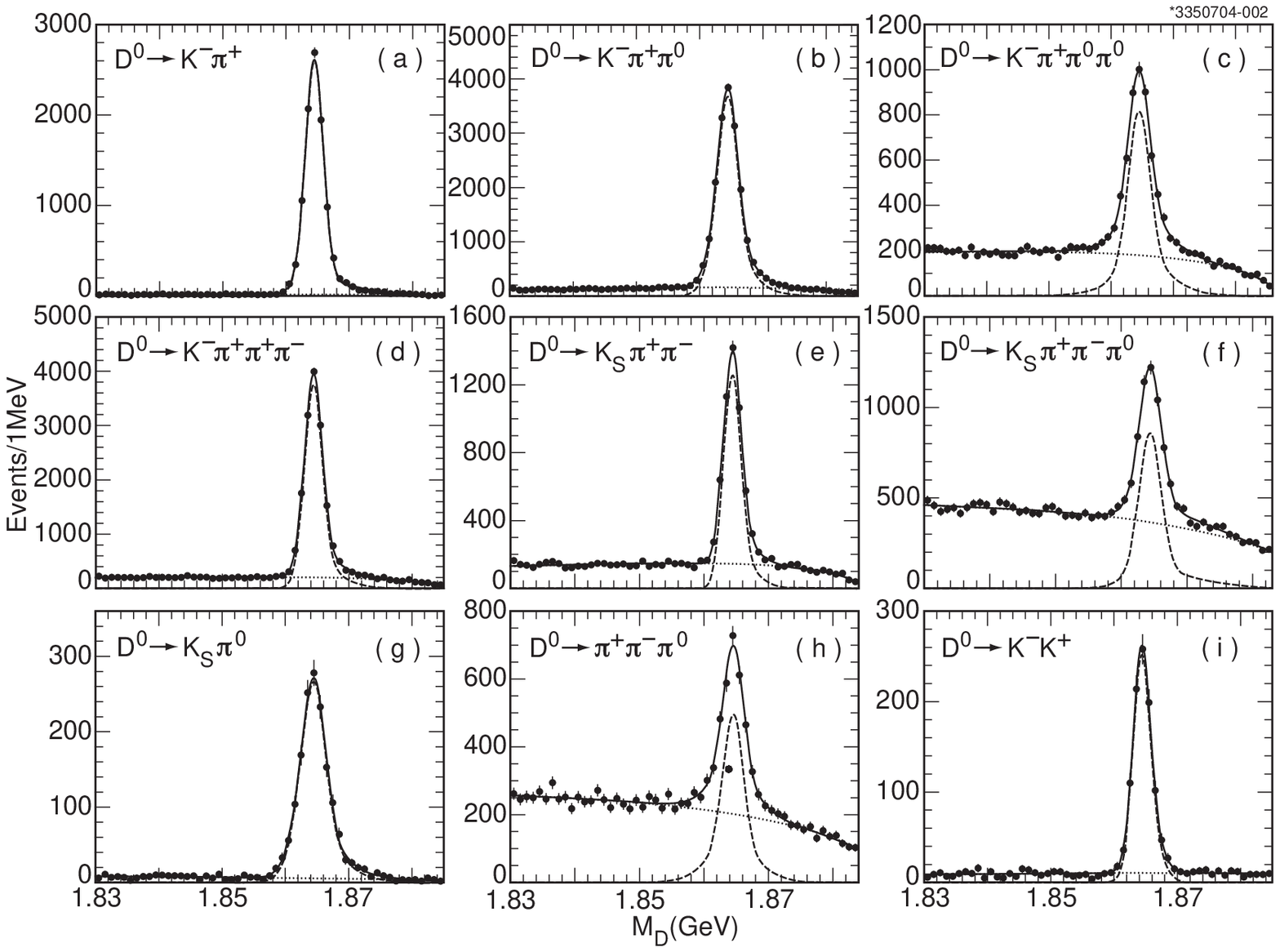,width=0.9\textwidth}} 
\vspace*{8pt}
\caption{Fits to the beam-constrained masses for different 
         fully reconstructed $D^{0}$ decay modes.
         The signal is described by a Gaussian and a bifurcated 
         Gaussian to account for the initial state radiation.
         The background is described by an Argus
         function $^3$.} 
\label{D0Tag} 
\end{figure}

We select events with a fully reconstructed $D^0$ meson where 
$D^0\to K^-\pi^+$,           $K^-\pi^+\pi^0$, $K^-\pi^+\pi^0\pi^0$, 
$K^-\pi^+\pi^+\pi^-$, $K_S\pi^0$,      $K_S\pi^+\pi^-$, 
$K_S\pi^+\pi^-\pi^0$, $\pi^+\pi^-\pi^0$, and $K^-K^+$. 
Charge conjugate decays are implied throughout this paper. 
Within the tagged events, we select the subset in which 
the $\bar D^0$ meson semileptonically decays to a specific 
final state. The efficieny-corrected ratio of the event yields
gives the absolute branching fraction for the exclusive semileptonic 
decay mode. The selection of the tag $D^0$ candidates is based on two 
variables $\Delta E=E_D-E_{beam}$ 
(the difference between the energy of the tag $D^0$ candidate ($E_D$)  and 
the beam energy ($E_{beam}$) ), and the beam constrained mass 
$M_{D}=\sqrt{E_{beam}^2-p_D^2}$, where $p_D$ is the momentum of the 
tag $D^0$ candidate. We use RICH and $dE/dx$ information to identify kaons 
and pions.  In Fig.~\ref{D0Tag}, we present the fits to the beam  
constrained masses of different tag modes. Multiple combinations have 
been eliminated by selecting the candidate with the minimum 
value of $|\Delta E|$. We have found about 60,000 tag $D^0$. 

\begin{figure}[hbtp]
\centerline{\psfig{file=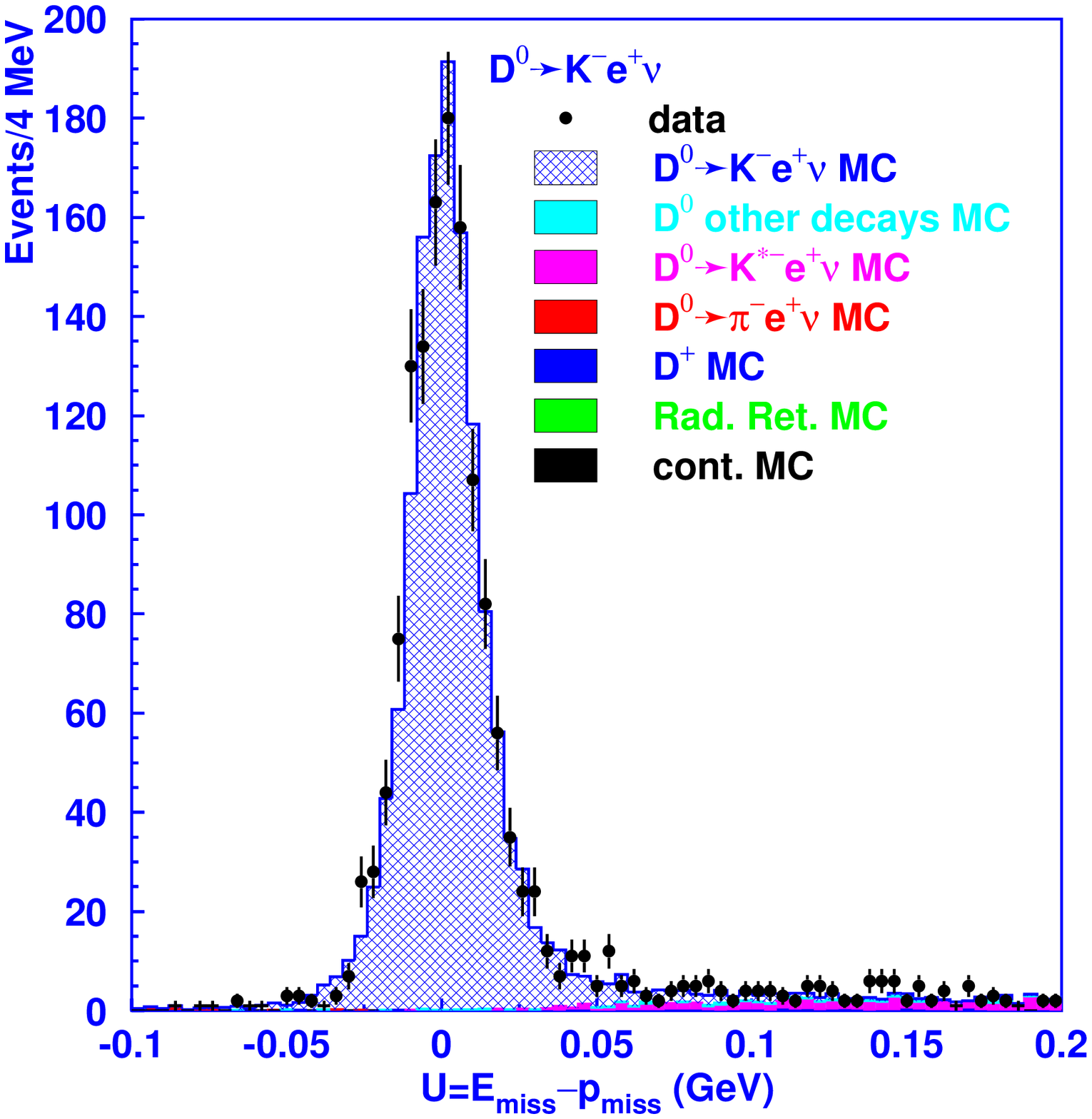,width=0.45\textwidth}
            \psfig{file=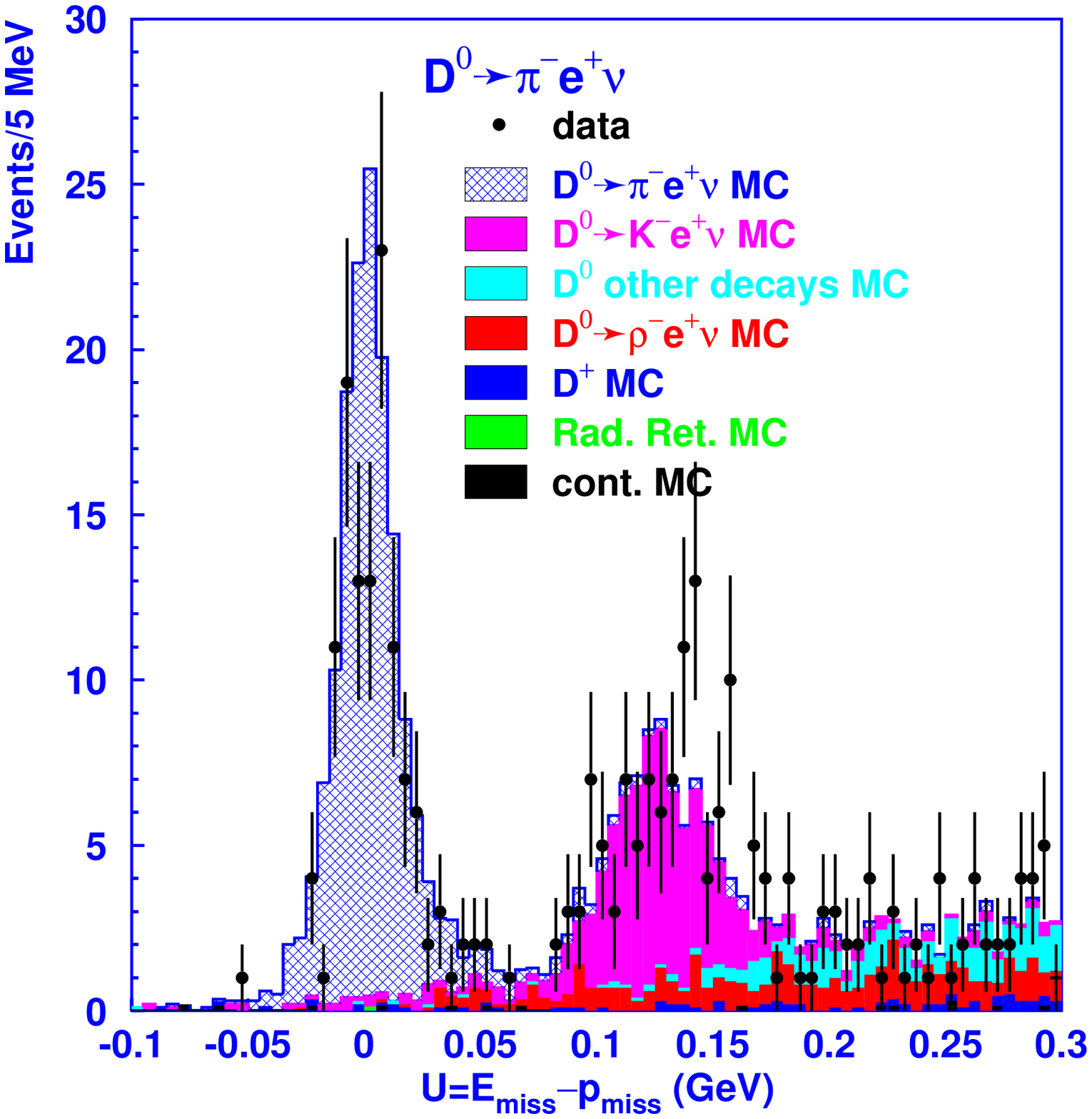,width=0.45\textwidth}}
\centerline{\psfig{file=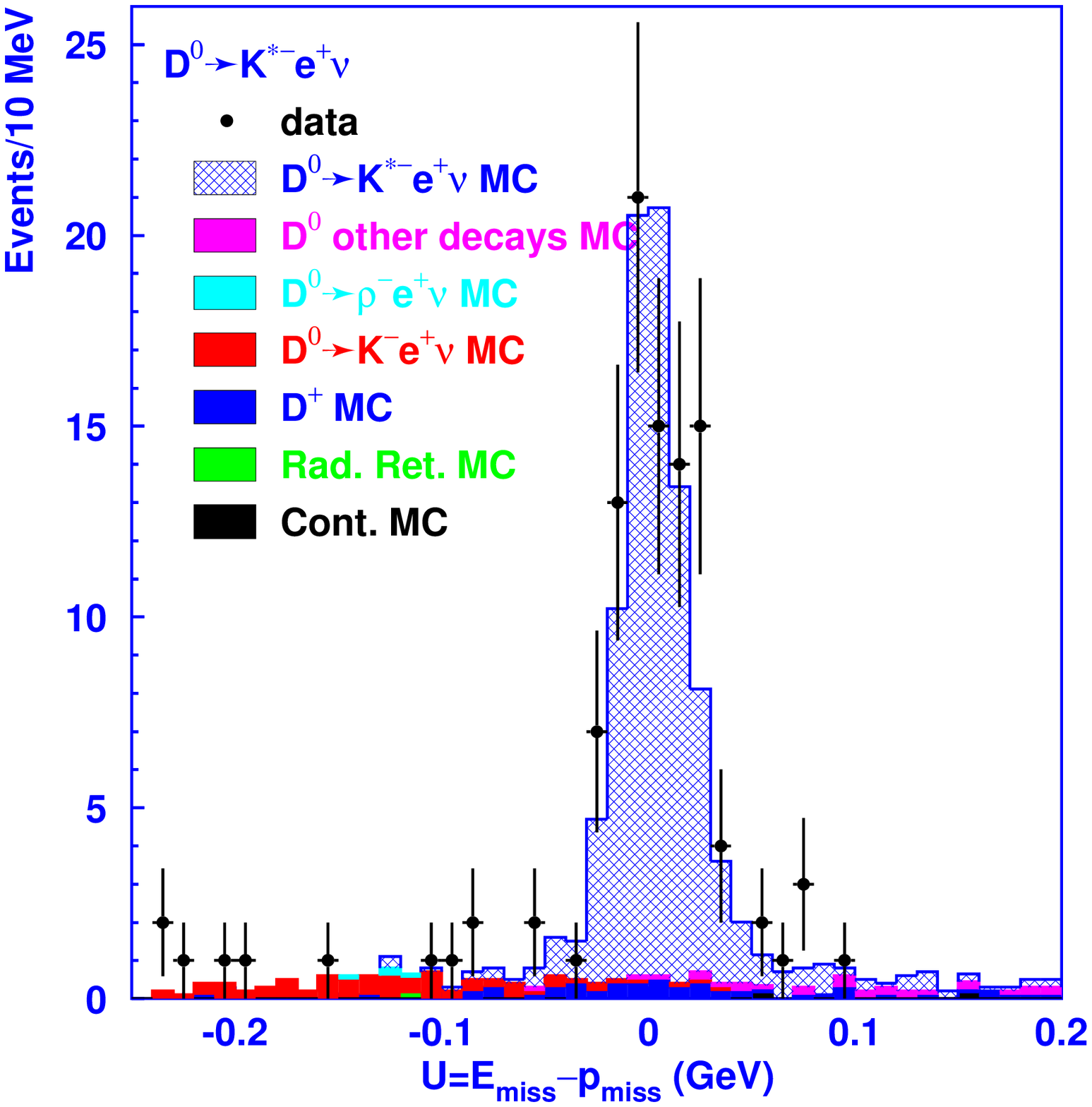,width=0.45\textwidth}
            \psfig{file=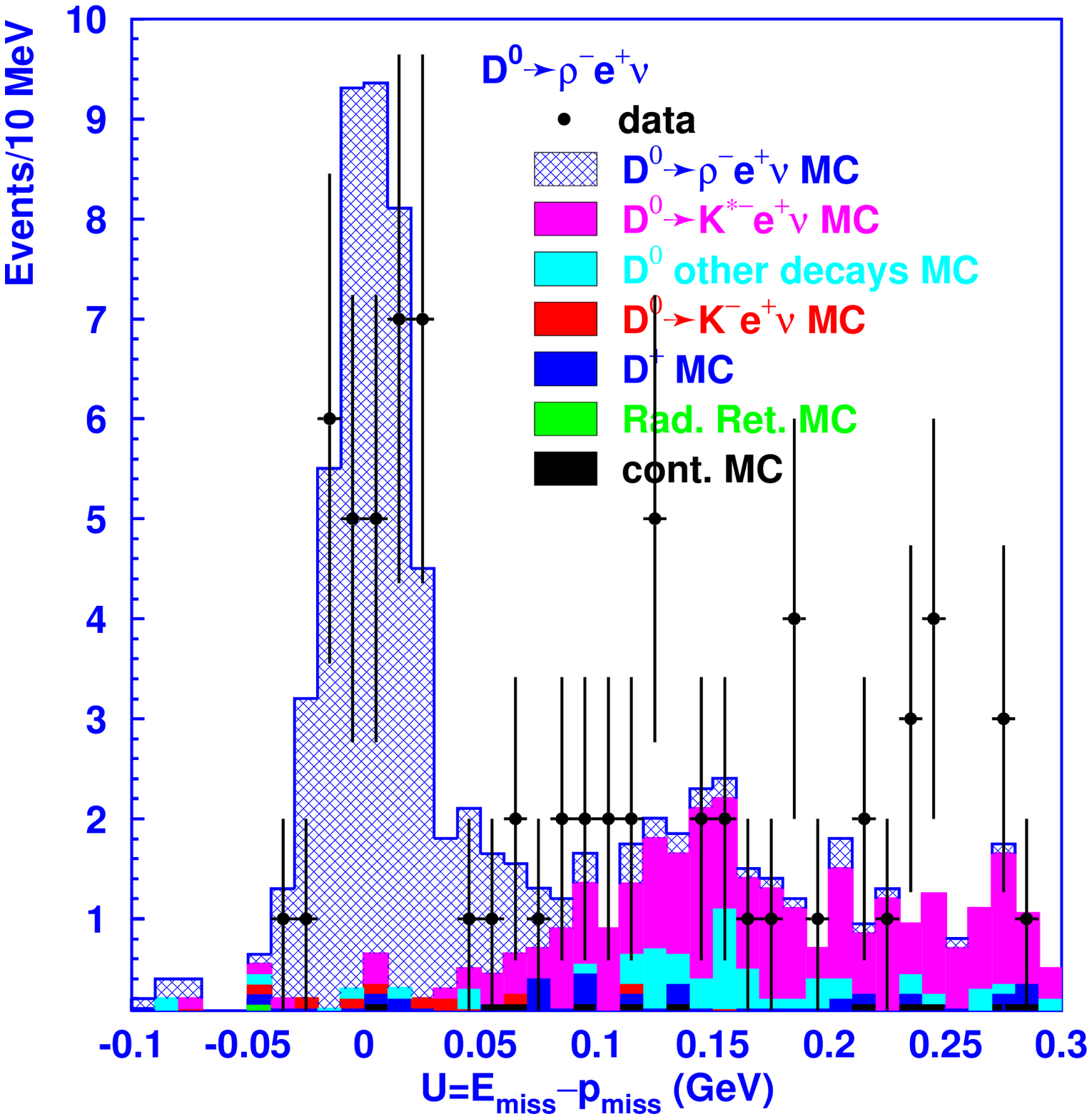,width=0.45\textwidth}}
\caption{Comparison of $U=E_{miss}-p_{miss}$ for the selected 
$D^0$ semileptonic decays between data and MC.} 
\label{DataMC} 
\end{figure}

We find $D^0$ semileptonic decays into $\ke$, $\pie$, $\kste$ 
($K^{*-}\to K^-\pi^0$) and $\rhoe$ ($\rho^-\to\pi^-\pi^0$) 
against a tag $D^0$ by reconstructing the difference of the missing 
energy and missing momentum which should peak at zero.  
In Fig.~\ref{DataMC}, we present the comparision for the selected 
$D^0$ semileptonic decay events between data and MC, the 
comparison shows good agreement. 

\begin{figure}[hbtp] 
\centerline{\psfig{file=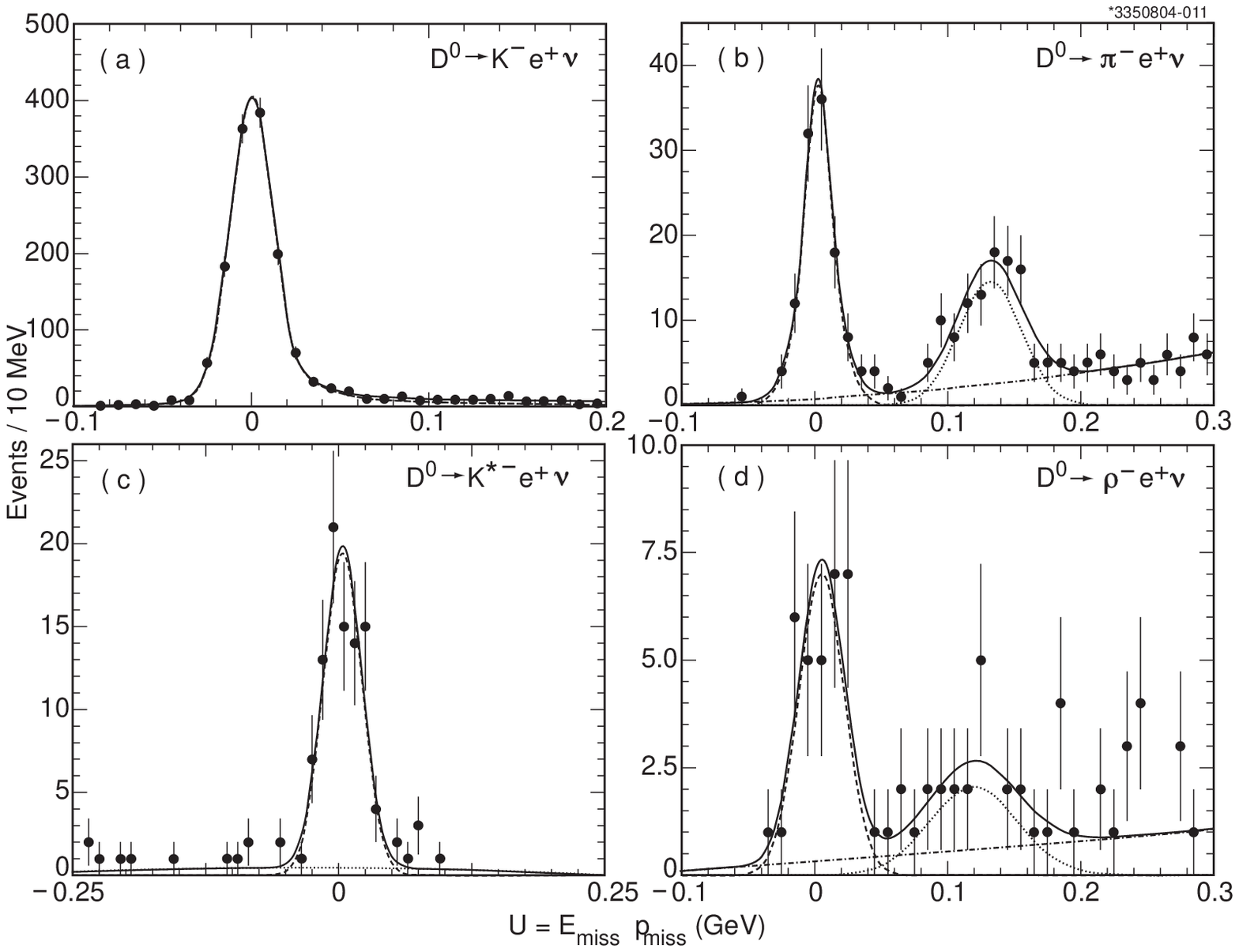,width=0.90\textwidth}}
\vspace*{8pt}
\caption{Fits to $U=E_{\rm miss}-p_{\rm miss}$ distributions for 
         $D^0 \to \ke$,  $\pie$, $\kste$ and $\rhoe$, with the other
         $\bar{D^{0}}$ fully reconstructed.} 
\label{semi} 
\end{figure}

The fits to the $U=E_{miss}-p_{miss}$ distributions 
 are shown in Fig.~\ref{semi}, and 
the yields are given in Table~\ref{summary}. 
The detail of event selection can be found in Ref.~\cite{Dhev}. 

\section{Results}  
The measured absolute branching fractions are given in Table~\ref{summary}, 
in comparison with the PDG values~\cite{PDG}. All results are preliminary. 
The errors are statistical and systematic, respectively. The dominant
systematic error comes from uncertainties of 
track and $\pi^0$ reconstruction efficiency
(3\% per track and 4.4\% per $\pi^0$)
which will improve with more data sample and further study. 

Based on a data sample of 60 pb$^{-1}$ collected at the $\psi(3770)$ resonance,
we have improved the absolute measurements of $D^0$ semileptonic decay
 branching fractions and presented the first observation of $\drhoe$.
Our results for $\dke$ and $\dkste$ are  consistent with those from the PDG. 
Our result ${\cal B}(\dpie)$ is lower than the PDG value. The 
ratio {${\cal B}(\dpie)\over{\cal B}(\dke)$} is close to the CLEO III result
$(8.2\pm0.6\pm0.5)\%$~\cite{CLEOIII}, while lower than the PDG value.

\begin{table}[htbp]
\tbl{Absolute branching fraction measurements of the exclusive 
         $D^0$ semileptonic decays, in comparison with the PDG. ~
         The uncertaities are statistical and systematic, respectively.
         }  
{\begin{tabular}{@{}cccc@{}} \toprule 
 Decays    & yields      & ${\cal B}$  & PDG  \\ \colrule 
$\dke$     & 1405.1$\pm38.5$ & (3.52$\pm0.10\pm0.25)\%$  & $(3.58\pm0.18)\%$ \\ 
$\dpie$    & 109.1$\pm$10.9  & (0.25$\pm0.03\pm0.02)\%$  & $(0.36\pm0.06)\%$ \\ 
$\dkste$   & 88.0$\pm$9.7    & (2.07$\pm0.23\pm0.18)\%$  & $(2.15\pm0.35)\%$ \\ 
$\drhoe$   & 30.1$\pm$5.8    & (0.19$\pm0.04\pm0.02)\%$  & none \\  \botrule 
{${\cal B}(\dpie)\over{\cal B}(\dke)$} & & $(7.0\pm0.7\pm0.3)\%$ 
           & $(10.1\pm1.8)\%$ \\ 
{${\cal B}(\drhoe)\over{\cal B}(\dkste)$} & & $(9.2\pm2.0\pm0.8)\%$ 
           & none \\ \botrule 
\end{tabular}} 
\label{summary}
\end{table}

In the near future, we expect to provide precision measurements of 
the absolute branching fractions for $D$ semileptonic decays at 
a precision of 2\%. While the normalizations of form factors 
$f_+(0)$ determined from Lattice QCD~\cite{quenched,unquenched} have  
10\% uncertainties. To determine $\vcs$ and $\vcd$ precisely, 
more precision Lattice 
QCD results are essential in the near future.

\section{Summary}
We have improved the absolute branching fractions for exclusive $D^0$ 
semileptonic decays. 
We expect more precise measurements of the absolute branching fractions 
for exclusive $D^0$ semileptonic decays in the near future, and 
therefore precision determinations of $\vcs$ and $\vcd$. 

We gratefully acknowledge the effort of the CESR staff in providing 
us with excellent luminosity and running conditions.
This work was supported by the National Science Foundation,
the U.S. Department of Energy, the Research Corporation,
and the Texas Advanced Research Program.

\end{document}